\def\aap #1 #2 #3 {{\sl Adv. Appl. Prob.} {\bf #1}, #2 (#3)}
\def\acp #1 #2 #3 {{\sl Adv. Chem. Phys.} {\bf #1}, #2 (#3)}
\def\ah #1 #2 #3 {{\sl Adv. Hydrosci.} {\bf #1}, #2 (#3)}
\def\aj #1 #2 #3 {{\sl Astrophys. J.} {\bf #1}, #2 (#3)}
\def\anp #1 #2 #3 {{\sl Ann. Phys.} {\bf #1}, #2 (#3)}
\def\ap #1 #2 #3 {{\sl Adv. Phys.} {\bf #1}, #2 (#3)}
\def\arpc #1 #2 #3 {{\sl Ann. Rev. Phys. Chem.} {\bf #1}, #2 (#3)}
\def\ast #1 #2 #3 {{\sl Aerosol Sci. Tech.} {\bf #1}, #2 (#3)}
\def\bj #1 #2 #3 {{\sl Biophys. J.} {\bf #1}, #2 (#3)}
\def\ces #1 #2 #3 {{\sl Chem. Engr. Sci.} {\bf #1}, #2 (#3)}
\def\cmc #1 #2 #3 {{\sl Chem. Engr. Comm.} {\bf #1}, #2 (#3)}
\def\eul #1 #2 #3 {{\sl Europhys. Lett.} {\bf #1}, #2 (#3)}
\def\ijf #1 #2 #3 {{\sl Int. J. of Fracture} {\bf #1}, #2 (#3)}
\def\ijmpb #1 #2 #3 {{\sl Int. J. Mod. Phys. B} {\bf #1}, #2 (#3)}
\def\ijss #1 #2 #3 {{\sl Int. J. Solids Structures} {\bf #1}, #2 (#3)}
\def\jap #1 #2 #3 {{\sl J. Appl. Phys.} {\bf #1}, #2 (#3)}
\def\japr #1 #2 #3 {{\sl J. Appl. Prob.} {\bf #1}, #2 (#3)}
\def\jas #1 #2 #3 {{\sl J. Atmos. Sci.} {\bf #1}, #2 (#3)}
\def\jcp #1 #2 #3 {{\sl J. Chem. Phys.} {\bf #1}, #2 (#3)}
\def\jcis #1 #2 #3 {{\sl J. Colloid Interface Sci.} {\bf #1}, #2 (#3)}
\def\jdep #1 #2 #3 {{\sl J. de Physique} {\bf #1}, #2 (#3)}
\def\jdepl #1 #2 #3 {{\sl J. de Physique Lett.} {\bf #1}, #2 (#3)}
\def\jetp #1 #2 #3 {{\sl Sov. Phys. JETP} {\bf #1}, #2 (#3)}
\def\jetpl #1 #2 #3 {{\sl Sov. Phys. JETP Letters} {\bf #1}, #2 (#3)}
\def\sps #1 #2 #3 {{\sl Sov. Phys. Semicond.} {\bf #1}, #2 (#3)}
\def\jfm #1 #2 #3 {{\sl J. Fluid Mech.} {\bf #1}, #2 (#3)}
\def\jgr #1 #2 #3 {{\sl J. Geophys. Res.} {\bf #1}, #2 (#3)}
\def\jkps #1 #2 #3 {{\sl J. Korean Phys. Soc.} {\bf #1}, #2 (#3)}
\def\jmp #1 #2 #3 {{\sl J. Math. Phys.} {\bf #1}, #2 (#3)}
\def\jms #1 #2 #3 {{\sl J. Memb. Sci.} {\bf #1}, #2 (#3)}
\def\jpa #1 #2 #3 {{\sl J. Phys. A} {\bf #1}, #2 (#3)}
\def\jpc #1 #2 #3 {{\sl J. Phys. C} {\bf #1}, #2 (#3)}
\def\jdep #1 #2 #3 {{\sl J. de Physique} {\bf #1}, #2 (#3)}
\def\jdepi #1 #2 #3 {{\sl J. de Physique I} {\bf #1}, #2 (#3)}
\def\jsp #1 #2 #3 {{\sl J. Stat. Phys.} {\bf #1}, #2 (#3)}
\def\jtb #1 #2 #3 {{\sl J. Theor. Biol.} {\bf #1}, #2 (#3)}
\def\jjpn #1 #2 #3 {{\sl J. Phys. Soc. Japan Suppl.} {\bf #1}, #2 (#3)}
\def\macro #1 #2 #3 {{\sl Macromolecules} {\bf #1}, #2 (#3)}
\def\nat #1 #2 #3 {{\sl Nature} {\bf #1}, #2 (#3)}
\def\nc #1 #2 #3 {{\sl Nuovo Cimento} {\bf #1}, #2 (#3)}
\def\npa #1 #2 #3 {{\sl Nucl. Phys. A} {\bf #1}, #2 (#3)}
\def\npb #1 #2 #3 {{\sl Nucl. Phys. B} {\bf #1}, #2 (#3)}
\def\pa #1 #2 #3 {{\sl Physica A} {\bf #1}, #2 (#3)}
\def\pla #1 #2 #3 {{\sl Phys. Lett. A} {\bf #1}, #2 (#3)}
\def\pf #1 #2 #3 {{\sl Phys. Fluids} {\bf #1}, #2 (#3)}
\def\pr #1 #2 #3 {{\sl Phys. Rev.} {\bf #1}, #2 (#3)}
\def\pra #1 #2 #3 {{\sl Phys. Rev. A} {\bf #1}, #2 (#3)}
\def\prb #1 #2 #3 {{\sl Phys. Rev. B} {\bf #1}, #2 (#3)}
\def\prc #1 #2 #3 {{\sl Phys. Rev. C} {\bf #1}, #2 (#3)}
\def\prd #1 #2 #3 {{\sl Phys. Rev. D} {\bf #1}, #2 (#3)}
\def\prl #1 #2 #3 {{\sl Phys. Rev. Lett.} {\bf #1}, #2 (#3)}
\def\prept #1 #2 #3 {{\sl Phys. Repts.} {\bf #1}, #2 (#3)}
\def\pt #1 #2 #3 {{\sl Powder Tech.} {\bf #1}, #2 (#3)}
\def\ptp #1 #2 #3 {{\sl Prog. Theor. Phys.} {\bf #1}, #2 (#3)}
\def\pams #1 #2 #3 {{\sl Proc. Am. Math. Soc.} {\bf #1}, #2 (#3)}
\def\pcps #1 #2 #3 {{\sl Proc. Camb. Philos. Soc.} {\bf #1}, #2 (#3)}
\def\pnas #1 #2 #3 {{\sl Proc. Natl. Acad. Sci.} {\bf #1}, #2 (#3)}
\def\prsl #1 #2 #3 {{\sl Proc. Roy. Soc. London Ser. A} {\bf #1}, #2 (#3)}
\def\rmp #1 #2 #3 {{\sl Rev. Mod. Phys.} {\bf #1}, #2 (#3)}
\def\sci #1 #2 #3 {{\sl Science} {\bf #1}, #2 (#3)}
\def\ssc #1 #2 #3 {{\sl Sol. State. Commun.} {\bf #1}, #2 (#3)}
\def\sur #1 #2 #3 {{\sl Surf. Sci.} {\bf #1}, #2 (#3)}
\def\wrr #1 #2 #3 {{\sl Water Resources Res.} {\bf #1}, #2 (#3)}
\def\ibid #1 #2 #3 {{\it ibid} {\bf #1}, #2 (#3)}
 
\documentstyle[12pt]{article}
\newcommand{\be}{\begin{equation}}
\newcommand{\ee}{\end{equation}}
\newcommand{\lpa}{\lambda_{\parallel}}  
\newcommand{\lpe}{\lambda_{\perp}}
\newcommand{\cpa}{C_{\parallel}(r)}  
\newcommand{\cpe}{C_{\perp}(r)}
\begin{document}
\baselineskip 0.8cm
\begin{center}
{\LARGE\bf Avalanche Size Distribution in \\ the Toom Interface \\} 
\vspace{1.0cm}
{H.~Jeong$^{\dag}$, B.~Kahng$^{\ddag}$, and D.~Kim$^{\dag}$ \\}
\vspace{1.0cm}
{\sl \dag Center for Theoretical Physics and Department of Physics, \\ 
Seoul National University, Seoul 151-742, Korea \\}
\vspace{0.5cm}
{\sl \ddag Department of Physics, Kon-Kuk University, Seoul 133-701, Korea \\}
\vspace{1.0cm}
{\Large Abstract}
\end{center}
We present numerical data of the height-height correlation function and 
of the avalanche size distribution for the three dimensional 
Toom interface. The height-height correlation function behaves samely 
as the interfacial fluctuation width, which diverges logarithmically 
with space and time for both unbiased and biased cases.
The avalanche size defined by the number of changing sites caused 
by a single noise 
process, exhibits an exponentially decaying distribution, which is in contrast 
to power-law distributions appearing in typical self-organized 
critical phenomena. We also generalize the Toom model into arbitrary dimensions.\\   
\vspace{1.0cm}

\vspace{2.0cm}

The Toom model [1] is a dynamical model of Ising spins in 
which the condition of detailed balance is not satisfied and 
hence whose phases are not described by equilibrium Gibbs ensembles. Recently,
Derrida, Lebowitz, Speer and Spohn (DLSS) [2] studied physical properties of
interfaces formed in the two-dimensional Toom model.
In low-noise limit, this model leads 
to a ($1+1$) dimensional solid-on-solid-type (SOS) model, 
which is in turn much simpler for understanding 
generic nature of dynamics.  
In the SOS model, the dynamics of spin-flips may be regarded as 
a deposition-evaporation process of particles. Due to the 
nature of the Toom dynamics, the deposition-evaporation process occurs in 
an avalanche fashion with preferred direction. 
DLSS found, among others, that the continuum stochastic equation describing
such SOS model is the well-known Kardar-Parisi-Zhang (KPZ) equation [3]. 
In reference [4],
present authors proposed a natural generalization of the model in three 
dimension on the body centered cubic lattice and found that the continuum 
equation is the anisotropic KPZ (AKPZ) equation given by 
\be
\partial_t h=\nu_{\parallel}{{\partial}_{\parallel}}^2 h 
+ \nu_{\perp}{{\partial}_{\perp}}^2 h 
+{1\over 2}\lpa(\partial_{\parallel}h)^2
+{1\over 2}\lpe(\partial_{\perp}h)^2+\eta
\ee
with {\it opposite signs} of the coefficients $\lpa$ and $\lpe$. 
In Eq.~(1), $h$ is the height of the fluctuating surface with respect 
to a reference (co-moving) plane, $\partial_{\parallel}$ ($\partial_{\perp}$) 
stands
for spatial derivative along the direction parallel (perpendicular) to the 
avalanche direction, and $\eta$ is the white noise. AKPZ equation with 
opposite signs of $\partial_{\parallel}$ and $\partial_{\perp}$ is known to 
renormalize toward the weak coupling limit [5] and consequently, square 
of the width, $w^2$, of the
Toom interface shows logarithmic dependence both in space and time. This is in
contrast to the model studied by Barab\'asi, Araujo, and Stanley [6] 
which again is described by Eq.~(1) but belongs to the strong coupling 
regime of the AKPZ universality.\\

In this work, we present numerical data of the height-height correlation 
function 
and of the avalanche size distribution for the three dimensional Toom 
interface. The height-height correlation function shows the same 
logarithmic dependences both on
 space and time as in $w^2$. The avalanche size distribution is found to 
be exponential. This is in contrast to power law behaviors appearing in 
typical self-organized critical phenomena and indirectly substantiate  
the validity of the collective variable approximation used by DLSS.\\

The $d$-dimensional Toom model we introduced in [4] and generalized here 
consists of
 Ising spins ($\sigma(x_1,x_2,\cdots,x_d)=\pm1$) on $d$-dimensional 
body-centered-cubic lattice. The spin coordinates $x_i$ takes the values 
in {\bf Z} for the
 spins on one sublattice and {\bf Z}+$\frac{1}{2}$ for those on the 
 other sublattice. At each time step, a randomly selected spin is updated 
according to
 the local rule that it becomes, at the next time step, equal to the majority
 of itself and of its $2^{d-1}$ neighbors relatively situated 
at $(-\frac{1}{2},\pm\frac{1}{2},\pm\frac{1}{2},\cdots,\pm\frac{1}{2})$ with 
probability $1-p-q$, to $+1$ with probability $p$, and to $-1$ with 
probability $q$. The referencing neighbors are shown in Fig.~1 for $d=3$.
 Unbiased (biased) dynamics results when $p=q$~$(p \neq q)$.
 To produce stationary interface at zero noise, the boundary spins on the 
three surfaces defined by $x_1=0$, $x_2=0$ and $x_2=L$, respectively, need 
to be fixed to the value $+1$~$(-1)$ if $x_2 >(<)$ $L/2$ where $L$ is the 
system size. Periodic boundary conditions are imposed to other surfaces.\\

In the low-noise limit, a spin flipped due 
to noise returns immediately to its original state by the majority rule, if
the spin is situated away from the interface.
Accordingly, we may assume that spin flips occur only at the interface. 
We are thus led to study an effective SOS-type
model as the generalization of the ($1+1$)-dimensional stairlike model studied 
by DLSS. In the SOS model picture, 
the Toom dynamics is mapped to a particle dynamics on 
a $d_s \equiv d-1$ dimensional substrate in the form of deposition-evaporation  
with avalanche. The ($d_s+1$) dimensional SOS model is defined on $d_s$ 
dimensional body-centered-cubic lattice. For $d_s=2$, it
is the checkerboard lattice, square lattice rotated by 45 degrees. To 
each lattice point, a relative height corresponding to the $x_2$ coordinate 
of the original Toom interface is assigned.
The height should satisfy the SOS condition, that nearest neighbor heights 
should differ only by $\pm1$.
Initially we begin with a flat surface characterized by height $0$ on one 
sublattice and height $1$ on the other. At each time step, we select a 
random site and start evaporation (deposition) process with 
probability ${\bar p}$ (probability $1-{\bar p}$).
In the evaporation (deposition) process, the height of the selected site is 
decreased (increased) by $2$ if the SOS condition is satisfied with respect 
to the $2^{d_s-1}$ spins to the left. By left spins, we mean those nearest 
neighbor spins whose $x_1$ coordinate is less by $\frac{1}{2}$. If the SOS 
condition to the left is not satisfied there is no change. Next the avalanche 
process proceeds to the right ($+x_1$ direction) until all sites satisfy the 
SOS condition.\\

When $d_s=1$, this avalanche dynamics reduces to the spin exchange dynamics 
of DLSS. The unbiased (biased) dynamics corresponds to the 
case ${\bar p} = \frac{1}{2}$~$ ({\bar p} \neq \frac{1}{2}$).
If the avalanche process is not allowed, so that 
depositions and evaporations occur only on local valleys and mountains 
respectively, then our model would be equivalent to 
the deposition-evaporation model proposed by Forrest and Tang [7], 
a generalization of the Plischke-R\'acz-Liu model [8] 
into higher dimensions. More detailed description of the SOS model for $d_s=2$ is 
found in the original paper [4]. Although the model is defined for 
general $d=d_s+1$, we confine our attention to $d=3$ from now on.\\

The continuum equation for the ($2+1$)-dimensional model is described by Eq.~(1), 
because it selects out a preferred direction, and because the 
cubic nonlinear term derived by DLSS was proved to be marginally 
irrelevant even in ($1+1$) dimensions [9]. 
For the unbiased case, both of the nonlinear terms in Eq.~(1) 
disappear due to the symmetry of deposition and evaporation. 
So the equation reduces to the Edwards-Wilkinson (EW) equation [10] implying that 
the square of the surface width diverges logarithmically with space 
and time. For the biased case, average height 
grows with increasing time, so that $\lpa$ and $\lpe$ are 
nonzero. It was shown [4] that $\lpa > 0$ and $\lpe < 0$ by applying 
the tilt argument [11]. 
Consequently, the model belongs to the weak-coupling 
regime of the AKPZ universality. Therefore, both the square of the surface 
width and the height-height correlation function are logarithmic 
for both unbiased and biased cases. \\ 

To check correspondences between the low noise Toom model and the SOS 
model, we performed numerical simulations for both models.
Results for both cases run 
on small sizes are in complete agreement with each other 
in low-noise limit. Accordingly, we performed 
simulations intensively for larger systems using the SOS model.  
The simulations are done in the range of system size 
$L=20\sim 140$ for both unbiased (${\bar p} = 0.5$) and 
biased (${\bar p} \neq 0.5$) cases.
We measured the surface width $w^2$ 
and the height-height correlation functions $C_{\parallel}(r_{\parallel},t)$ 
and $C_{\perp}(r_{\perp},t)$, defined by
\be
C_{\parallel}(r_{\parallel},t) =\langle[h(r_{\parallel},t)-h(0,0)]^2\rangle,
 \phantom{xx} \hbox{and} \phantom{xx} 
C_{\perp}(r_{\perp},t) =\langle[h(r_{\perp},t)-h(0,0)]^2\rangle, 
\ee
respectively. Here, $r_{\parallel}$~$(r_{\perp})$ denotes the spatial 
separation along the direction parallel (perpendicular) to the 
avalanche direction. 
From the theoretical considerations, we expect the height-height correlation 
functions diverges as $\sim \ln t$ before saturation, and $\sim \ln r$ after 
saturation.
Typical data of $\cpa$ and $\cpe$ after saturation are shown in Fig.~2 for 
the case of unbiased dynamics (${\bar p}=0.5$) and in Fig.~3 for the case of 
biased one (${\bar p}=0.3$). Each curve is for system 
sizes $L=$ 40, 60, 80, 100, and 120, respectively, from bottom to top and 
is averaged over 300 configurations. It confirms clearly the logarithmic 
dependence on $r$. Deviation from straight line for large $r$ is the 
finite size effect.\\

Next, we examined the avalanche size distribution 
$n(s)$. The avalanche size $s$ is defined as the number of successive spin flips 
by a single noise process. $n(s)$ was measured  
in two different manners. In the first case, it is measured in the 
critical state (after saturation), while in the second case, it is 
measured during the whole time steps. In both cases, 
the distribution function $n(s)$ is found to be exponential, 
$n(s)\sim \exp(-s/s^*)$ as shown in 
Figs.~4 and 5. The characteristic size $s^*$ determined from the slopes of
the inset figures is found to be independent of ${\bar p}$ and of the way
$n(s)$ is measured. It takes the value $s^*=1.11 \pm 0.01$.\\  

In conclusion, we have generalized the Toom model into $d$ dimensions and defined 
its associated SOS-type model. Also we have presented 
numerical data for the height-height correlation functions 
in perpendicular and parallel directions, respectively, for unbiased 
and biased cases for $d=3$.  
We have found that for the unbiased case, the interface is described 
by the EW equation, and for the biased case, it is described 
by the AKPZ equation with the opposite signs 
of $\lambda_{\parallel}$ and $\lambda_{\perp}$. 
Consequently, the height-height correlation functions diverge logarithmically 
with space and time. The avalanche size distribution has also been examined, 
which exhibits an exponential distribution for unbiased 
and biased cases instead of a power-law distribution. \\ 

We would like to thank Dr.~Jin~Min~Kim for helpful discussions. 
This work was supported in part by the Korea Science and Engineering
Foundation through the SRC 
program of SNU-CTP and through CTSP in Korea Univ, and in part  
by the Ministry of Education, Korea. \\

\vspace{1.0cm}

\vspace{1.0cm}
\newpage
{\Large\bf Figure Captions }
\begin{description}
\item[Fig.~1 ] The three dimensional Toom rule on bcc lattice used 
in this work. The black circled spin in the center is updated with 
the majority rule of itself and four nearest neighbor spins 
(the black circled) with probability $1-p-q$, and becomes equal to $+1$ ($-1$) 
with probability $p$ ($q$). In the low-noise limit, $p, q \rightarrow 0$.   
\item[Fig.~2 ] The height-height correlation functions 
$\cpa$ (dotted line) and $\cpe$ (solid line) versus $\ln r$ for unbiased case,
after saturation.
Each curve is for system sizes $L= 40, 60, 80, 100$ and $120$,
respectively,  from bottom to top, and is averaged over 300 configurations. 
\item[Fig.~3 ] Same as in Fig.~2 for biased case, ${\bar p}=0.3$.
\item[Fig.~4 ] The avalanche size distribution $n(s)$ versus $s$ for 
unbiased case measured during whole time 
steps (a) and after saturation (b). The data are obtained from different 
system size, but they are collapsed into each other, implying that $n(s)$ is 
independent of system size. Insets show $\ln n(s)$ versus $s$.
\item[Fig.~5 ] Same as in Fig.~4 for biased case, ${\bar p}=0.3$.
\end{description}
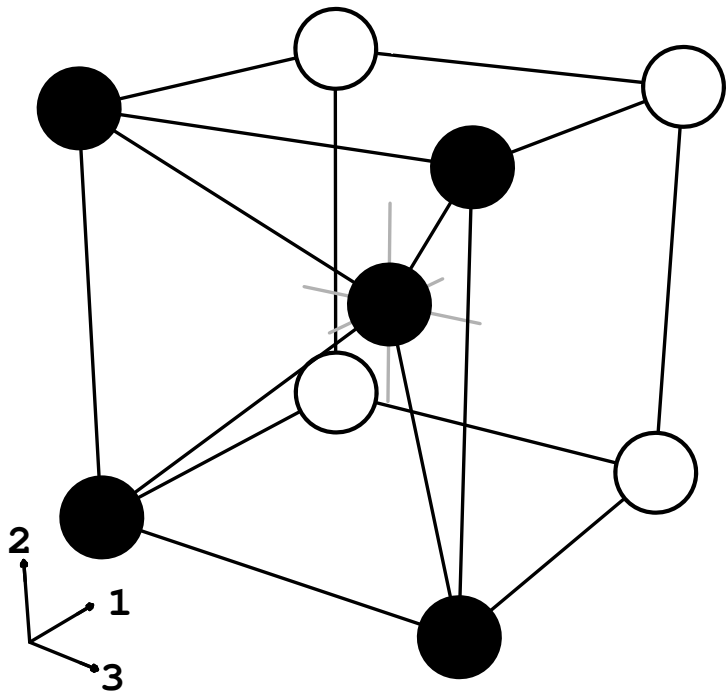
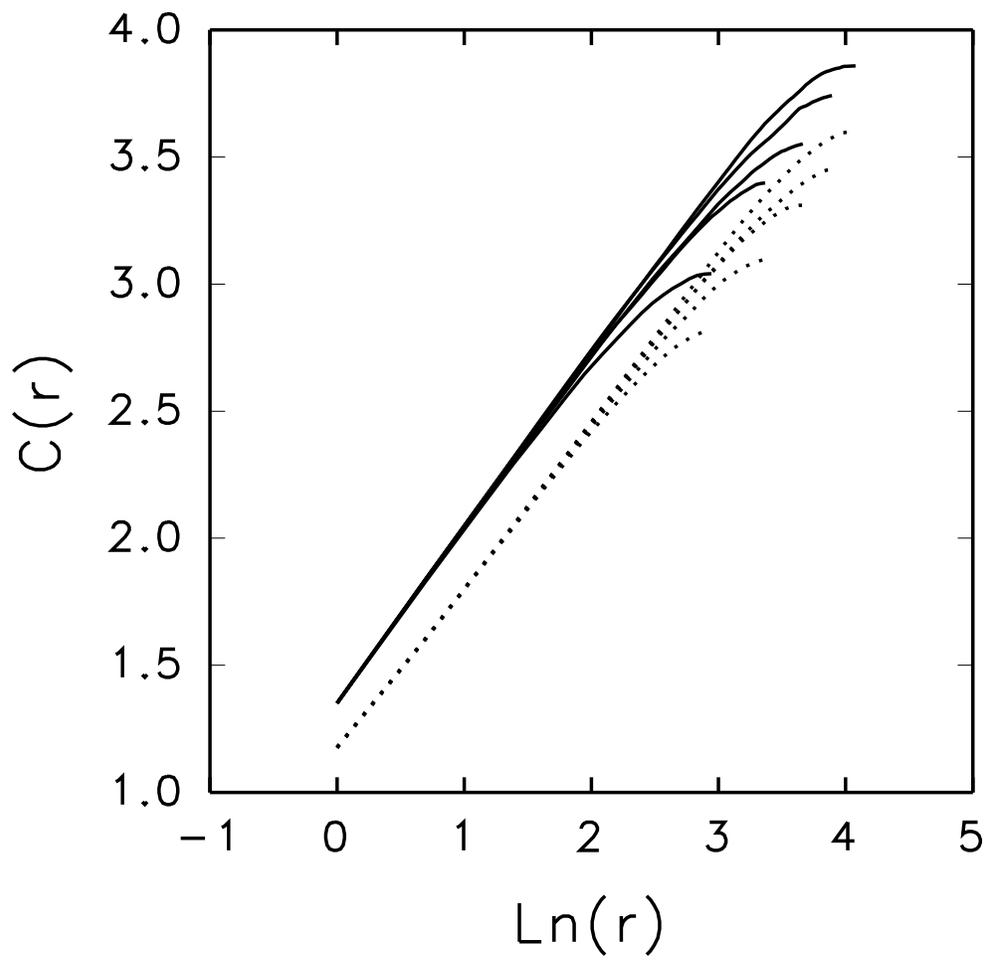
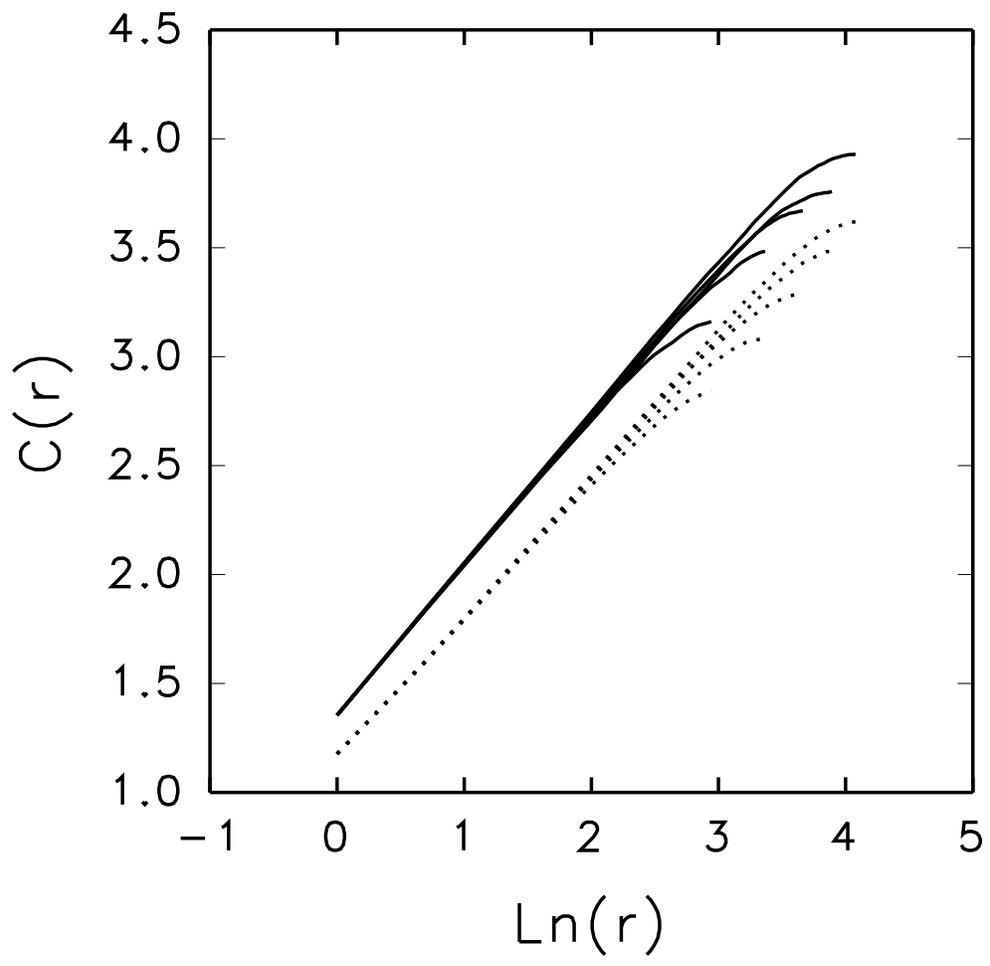
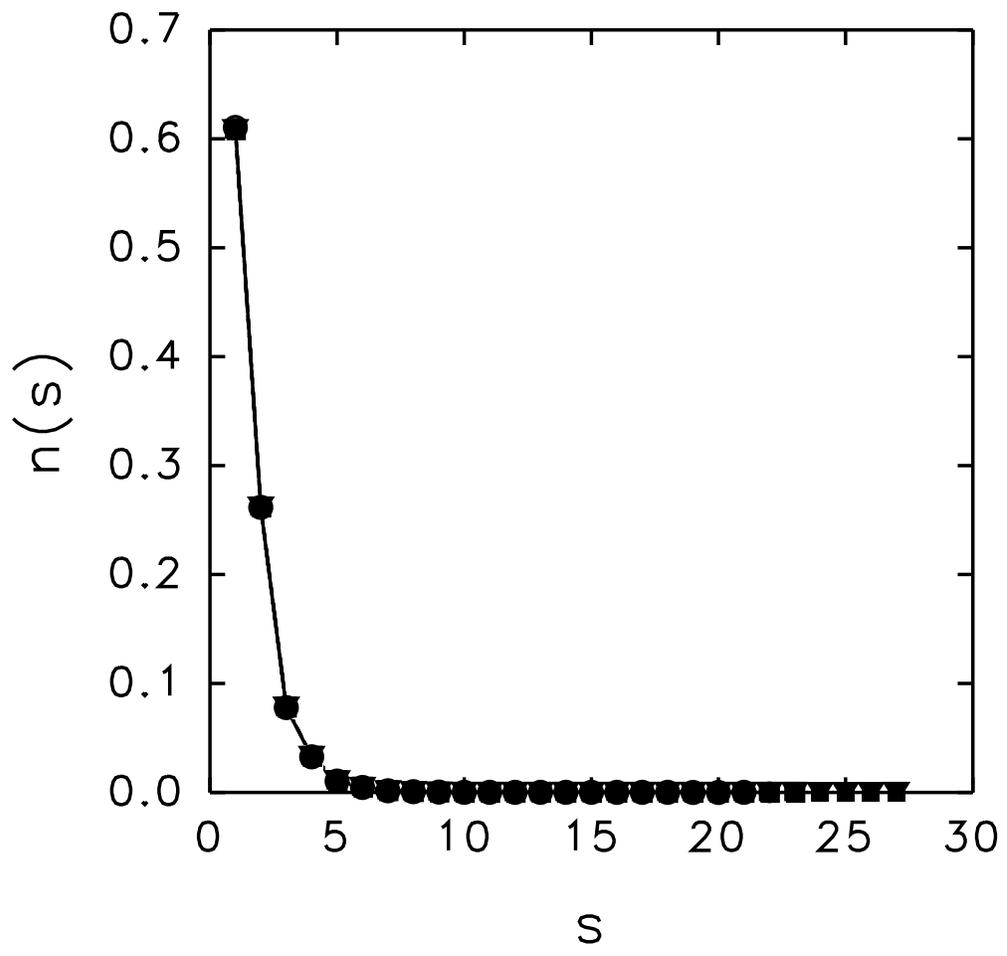
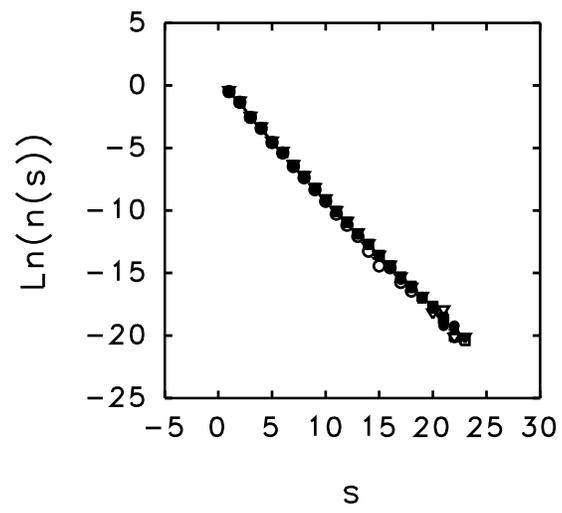
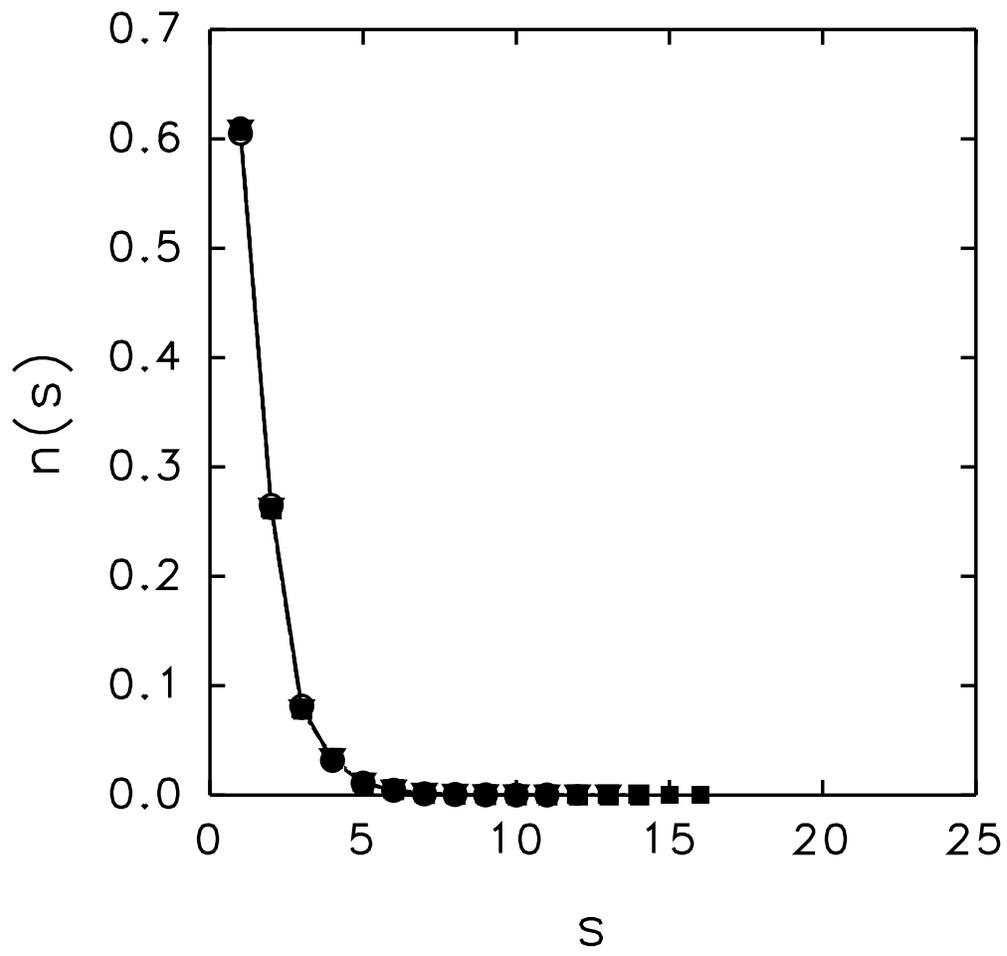
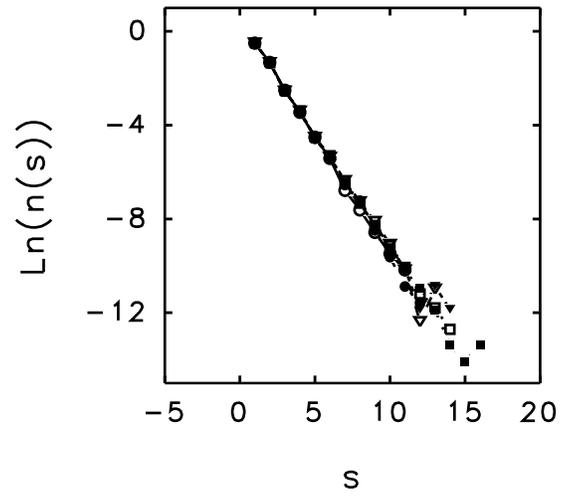
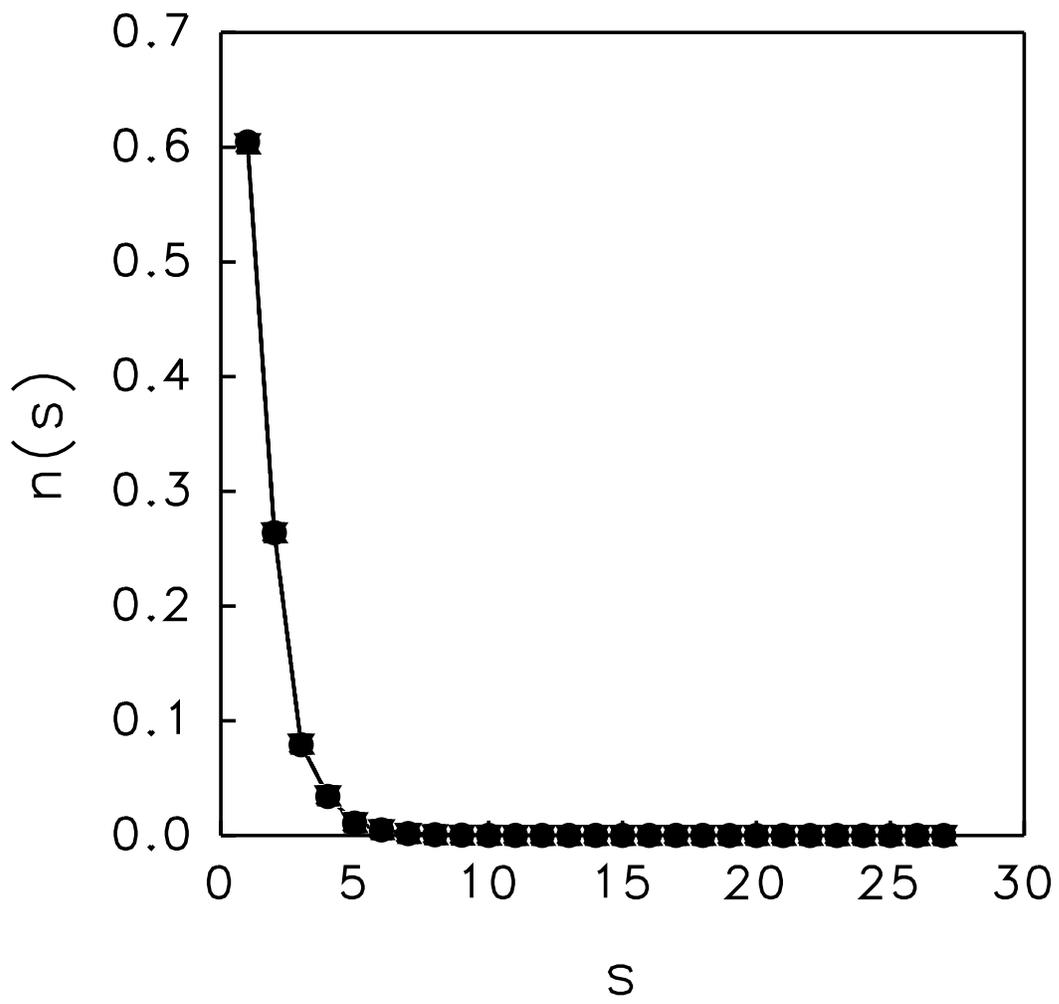
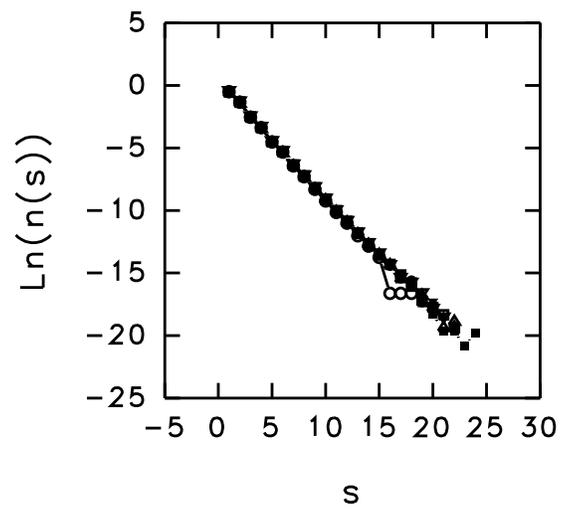
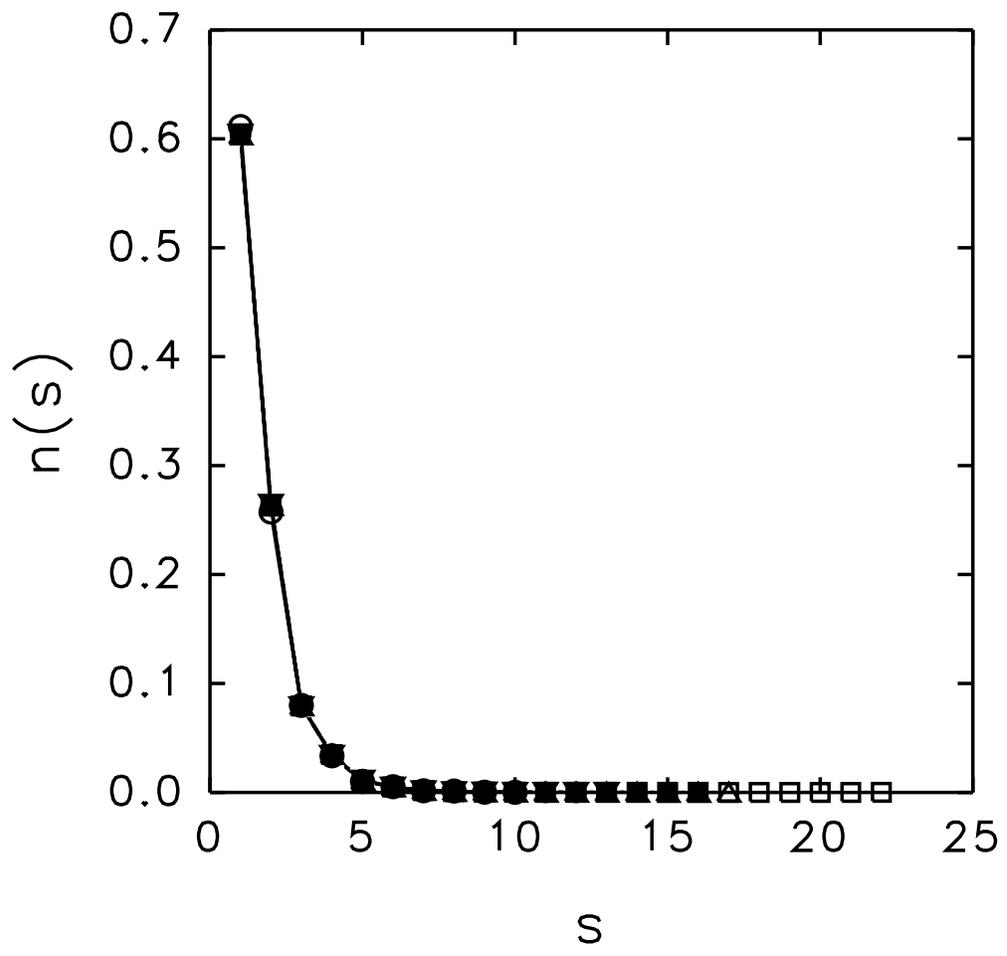
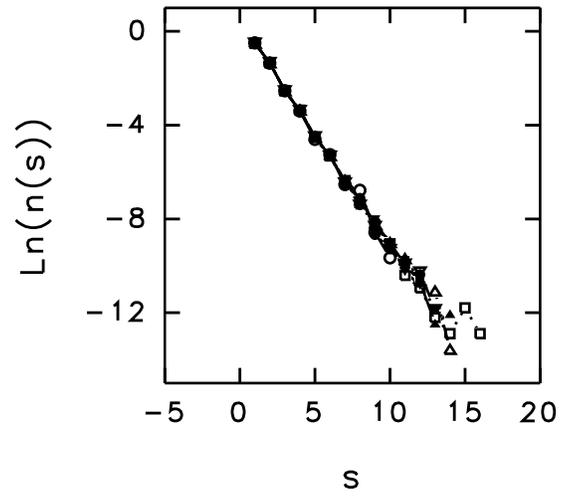
\end{document}